\documentclass[twocolumn,showpacs,preprintnumbers,amsmath,amssymb,pre]{revtex4}

\usepackage{graphicx}
\usepackage{dcolumn}
\usepackage{bm}
\usepackage{color}

\usepackage[english]{babel}
\usepackage[latin2]{inputenc}
\usepackage[T1]{fontenc}
\usepackage{ae,aecompl}

 \def\ccc#1;#2{\left\langle #1 \left\vert #2 \right.\right\rangle}
\def\ev #1{\left\langle #1 \right\rangle}

\begin{document}

\preprint{}

\title{Random walks on complex networks with inhomogeneous impact}

\author{Zolt\'an Eisler}
\email{eisler@maxwell.phy.bme.hu} \homepage{http://maxwell.phy.bme.hu/~eisler}
\author{J\'anos Kert\'esz}
\altaffiliation[Also at ]{Laboratory of Computational Engineering, Helsinki University of Technology, Espoo, Finland} \affiliation{Department of Theoretical Physics, Budapest University of Technology and Economics, Budapest, Hungary}

\date{\today}

\begin{abstract}
In many complex systems, for the activity $f_i$ of the constituents or
nodes $i$ a power-law relationship was discovered between the standard
deviation $\sigma_i$ and the average strength of the activity: $
\sigma_i \propto \left\langle f_i\right\rangle^\alpha $; universal
values $\alpha =1/2$ or $1$ were found, however, with exceptions. With
the help of an impact variable we introduce a random walk model where
the activity is the product of the number of visitors at a node and
their impact. If the impact depends strongly on the node connectivity
and the properties of the carrying network are broadly distributed
(like in a scale free network) we find both analytically and
numerically non-universal $\alpha$ values. The exponent always crosses
over to the universal value of $1$ if the external drive
dominates. \end{abstract}

\maketitle

Complex systems usually consist of many interacting units such that
their scaffold is a network. The most puzzling
questions are usually the ones regarding the dynamics of such
systems. Examples range from traffic (vehicular or Internet) through
the biochemistry of the cell to markets which are
systems generated by human interactions. Due to its generality, the
network aspect has turned out to be extremely fruitful in
these studies, but the dynamics is usually rather
individual, system dependent. However, recently an interesting
unifying feature was found in systems with multichannel
observations.

Processes taking place in many complex systems can be characterized by
generalized \emph{activities} $f_i(t)\geq 0$, defined at each
constituent or node $i=1\dots N$. For a wide range of systems,
research
\cite{barabasi.fluct,barabasi.separating,eisler.non-universality} has
revealed power-law scaling between the mean and the standard deviation
of the activity of the nodes:
\begin{equation} \sigma_i \propto \ev{f_i}^\alpha ,
\label{eq:power-law}
\end{equation}
where by definition
\begin{equation}
\sigma_i = \sqrt{\ev{\left ( f_i - \ev{f_i}\right )^2}}.
\end{equation}

This is not unmotivated from equilibrium statistical physics. Many
physical systems belong to the class $\alpha = 1/2$; in most cases
this is the fingerprint of equilibrium and the dominance of
\emph{internal dynamics}. Examples of such behavior are a
computer chip or the hardware level Internet (the network of data
transmission) \cite{barabasi.fluct}.

One can show analytically the existence of a universality class with
$\alpha = 1$. This value always prevails in the presence of a strong
driving force, when the dominant factor is such \emph{externally
imposed dynamics}. This limit is found for river networks, highway
traffic and the World Wide Web \cite{barabasi.fluct}.

It is instructive to recall that although only different aspects of
the same system, Internet and WWW fall into separate
categories. Former has a robust internal activity even without outside
(human) interaction, due to automatic queries, data transfer, etc. On
the other hand, latter consists of web pages whose activity is
generated by external demand (i.e., the clicks of users).

For such an analysis multichannel monitoring of a large number of
elements is needed with a possibly broad range of $\ev{f_i}$. A
previous work \cite{barabasi.separating} introduced a method of
decomposition to separate the effects of an external driving force
from the system's internal dynamics originating from the constituents'
individual behavior and interactions. Furthermore, when external
driving was absent or subordinate, $\alpha = 1/2$ seemed to hold for
every investigated system, though universality could not be proven. In
fact, more recent studies have shown that there are exceptions from
this rule: Fluctuations of stock market trading activity (traded
volume times price) are
characterized by $\alpha \approx 0.72$
\cite{eisler.non-universality}. In this paper we present a
generalization of a random walk model by Menezes and Barab\'asi
\cite{barabasi.fluct}, which accounts for such anomalous $\alpha$
values and clarifies their microscopic origins.

The model in Ref. \cite{barabasi.fluct} was motivated by the
statistics of web page visitations and we will also use this language.
Let us take a scale free Barab\'asi-Albert network of $N$ nodes (web
pages) \cite{barabasi.rmp, foot1}. First, we distribute $W$ tokens
(users, walkers) on the nodes randomly. In every time step every
walker jumps from its present site to a random neighbor of the site
along an edge (link). The new feature is that if a user steps to a
site, it has to pay a certain value $V(i)$ (exerts a certain impact)
which depends on the degree $k(i)$ of the visited node (more popular
pages with more links cost more money) as a power-law:
\begin{equation}
V(i) \propto k(i)^{\mu }.
\label{eq:vi}
\end{equation}
We must emphasize, that although this is a simplification, the
assumption is not ad hoc. E.g., the average activity for a given
stock was found to scale with the exponent $\mu \approx 0.44$ as a
function of the capitalization of the companies \cite{zumbach}, which,
in a sense, can be identified with the nodes' strength. This
application will be discussed in detail.

We continue the diffusion for $T$ steps and finally calculate the
total profit of every web site, which equals the $N(i)$ number of
visitations multiplied by $V(i)$; these values will be the 
activities $f_i(t=1)$ of the nodes on day $1$. We iterate this
procedure for the same network for $t=1, \dots , D$ days to generate
the whole time series $f_i(t)$. The original variant presented in
Ref. \cite{barabasi.fluct} corresponds to the special case $\mu = 0$:
it only counts $N(i)$ for each node. Later we will address the case
when the $V(i)$ values are allowed to fluctuate.

We expect a power-law scaling relationship such as
\begin{equation}
\ev{N(i)}\propto k(i)^\nu ,
\end{equation}
i.e., the mean number of the visitations to a node is proportional to
its degree. Higher connected vertices should have higher
traffic. Stationary solution of the master equation yields $\nu = 1$
analytically, in line with Ref. \cite{noh}. Also, if initially tokens
are distributed uniformly, after the first step the expectation value
of the number of tokens at any node will be directly proportional to
its degree. Therefore the model reaches the stationary state in one
step. For the sake of generality we will keep the notation for a
general value of $\nu$, which could be generated by different
dynamics.

We can write the mean profit of node $i$ as
\begin{equation}
\ev{f_i} = \ev{N(i)V(i)} \propto k(i)^{\mu+\nu}.
\label{eq:fi}
\end{equation}
Fluctuations can be expressed by the application of the central limit
theorem. As walkers do not interact, their visits are independent, and
thus for large enough $\ev{N(i)}$ and finite $\ev{N(i)^2}$ the
distribution converges to a Gaussian. Moreover, the variance of the
visits at node $i$ is $\sigma_{Ni}^2 = \ev{(N(i)-\ev{N(i)})^2} \propto
\ev{N(i)} \propto k(i)^\nu$. The variance of the signal detected on
node $i$ can then be written as
\begin{equation}
\sigma_i^2 = \sigma_{Ni}^2V(i)^2 \propto k(i)^{2\mu+\nu},
\label{eq:sigmai}
\end{equation}
where the proportionality comes from \eqref{eq:vi}. Finally, one can
combine \eqref{eq:fi} and \eqref{eq:sigmai} to get \begin{equation}
\alpha = \frac{1}{2}\left(1+\frac{\mu/\nu}{\mu/\nu+1}\right ).
\label{eq:alpha}
\end{equation}
$\alpha$ is the \emph{internal dynamical exponent} defined by
\eqref{eq:power-law} \emph{in the absence of external forces}.

\begin{figure}[tb]
\centerline{\includegraphics[width=185pt]{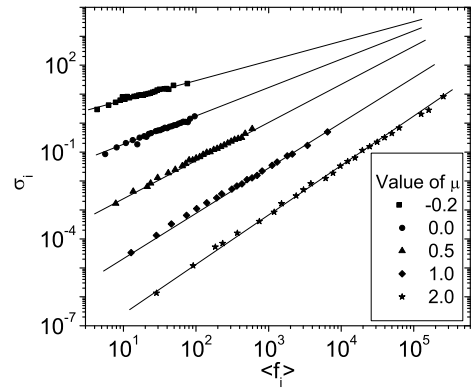}}
\caption{Scaling of the standard deviation of node activity with the
mean signal of the same node. A single point in the graph represents
the average standard deviation of all nodes with approximately the
same flux. Slopes on the log-log scale give the internal dynamical
exponent $\alpha$. Varying the impact distribution by changing $\mu$
causes a continuous change in $\alpha$, as expected from
\eqref{eq:alpha}.}
\label{fig:rwm_internal_scaling} \end{figure}

\begin{figure}[tb]
\centerline{\includegraphics[width=185pt]{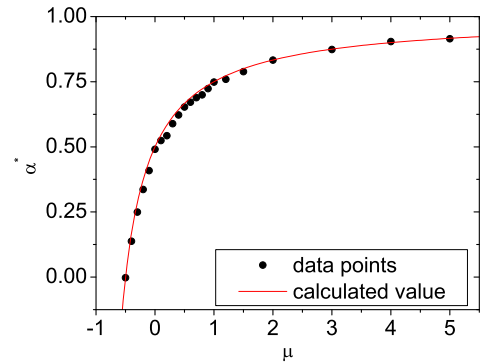}}
\caption{The values of the internal dynamical exponent $\alpha$ as a
function of $\mu$ governing the distribution of node dependent impacts. Circles
show simulation results ($N=2000$ nodes, $W=200$ walkers, $T=100$
steps/day, averaged for $D=10^5$ days). The solid line represents the
analytical formula \eqref{eq:alpha} for $\nu = 1$. By setting $\mu =
0$, we recover $\alpha=1/2$, which is observed for several equilibrium
systems. If one allows for node dependent impacts ($\mu \not = 0$),
non-universal behavior emerges and $\alpha$ can change continuously
between $0$ and $1$.} \label{fig:rwm_internal}
\end{figure}

We performed simulations of such a process and found perfect agreement
with the above calculation. We fixed a Barab\'asi-Albert network of
$N=2000$ nodes, $W=200$ tokens, $T=100$ steps per day, and averaged
over $D=10^5$ days. We also varied $\mu=-0.5,\dots,5.0$. Examples for
the scaling relation \eqref{eq:power-law} are shown in
Fig. \ref{fig:rwm_internal_scaling} \cite{foot2}. There is a clear
dependence of the slope on the value of $\mu$. The measured exponents
$\alpha$, compared with the analytical formula are shown in
Fig. \ref{fig:rwm_internal}.

The r.h.s. of \eqref{eq:alpha} is governed by the single parameter
$\mu/\nu$. By setting $\mu = 0$, we recover the original, non-independent
impacts and $\alpha = 1/2$. If $\mu/\nu > 0$, the scaling exponent of
fluctuations increases, $\alpha > 1/2$. As $\mu/\nu\rightarrow
\infty$, $\alpha \rightarrow 1$, which is the same exponent but due to
a different mechanism as $\alpha=1$ arising from strong driving. Note
that by choosing $\mu/\nu < 0$, $\alpha<1/2$ values are also accessible.

This result is robust against fluctuations in $V$, i.e., if different
users spend different amounts of money while visiting the same web
page, provided
\begin{equation}
V(i) = \ev{V(i)}\cdot X \propto k(i)^\mu \cdot X.
\label{eq:vinoisy}
\end{equation}
$X$ is drawn independently from a fixed distribution for every
visitation of node $i$. It is also assumed to have a finite second
moment.

The distributions of $N(i)$ and $V(i)$ are
independent and so they factorize in \eqref{eq:fi}, which formula
hence remains unchanged. In order to prove that scaling suggested by
\eqref{eq:sigmai} also persists, let us write
\begin{equation}
\sigma_i^2 = \ev{\left (\sum_{n = 1}^N V_i(n) - \ev{\sum_{n = 1}^N
V_i(n)}\right )^2},
\label{eq:proof1}
\end{equation}
where $n$ runs for all the $N$ visits to site $i$ during the day and
$V_i(n)$ is the profit from the $n$th visit.  By denoting
$\sum_{n=1}^N V_i(n)$ as $V_N$, its density function as $\mathbb
P(V_N)$, and that of $N(i)$ by $\mathbb P(N)$, it is possible to
rewrite \eqref{eq:proof1} as
\begin{eqnarray}
\sigma_i^2=\sum_{N=0}^\infty \mathbb P(N)\int_0^\infty dV_N\mathbb
P(V_N)V_N^2- \nonumber \\ \left (\sum_{N=0}^\infty \mathbb P(N)\int
dV_N\mathbb P(V_N)V_N\right )^2.
\label{eq:proof2}
\end{eqnarray}
By both adding and subtracting the term $\sum_{N=0}^\infty \mathbb
P(N)\left (\int_0^\infty dV_N\mathbb P(V_N)V_N\right )^2$, then
applying the equality $\int_0^\infty dV_N\mathbb P(V_N)V_N=N\ev{V(i)}$
and that for any fixed $N$ the variance of $V_N$ is
$\sigma^2_{V_N}=N\sigma^2_V$, one finally finds
\begin{equation}
\sigma_i^2=\sigma^2_{Ni}\ev{V(i)}^2+\sigma_{Vi}^2\ev{N(i)} \propto
k(i)^{2\mu+\nu}.
\label{eq:proof3}
\end{equation}
The final proportionality comes from similar arguments as in the case
of \eqref{eq:sigmai}. In particular, we know that $\sigma_{Ni}^2
\propto \ev{N(i)} \propto k(i)^\nu$. On the other hand, with respect
to scaling with the node degree $k(i)$, we defined $V(i)^2 \propto
k(i)^{2\mu }$, while also $\sigma^2_V \propto k(i)^{2\mu}$.

One can see explicitly the new source that contributes to
fluctuations. The first term, basically the same as before, comes
from diffusive dynamics. The second, additional term is which
describes the effect of visit to visit variations present in impacts
$V(i)$. Regardless of this more complicated structure, the
scaling of $\sigma_i^2$ with the vertex degree $k(i)$ is preserved,
similarly to $\ev{f_i}$. Thus the dynamical exponent $\alpha$ is
unaffected. Simulations based on various distributions of $X$
confirmed this calculation.

\begin{figure}[!t]
\centerline{\includegraphics[width=185pt]{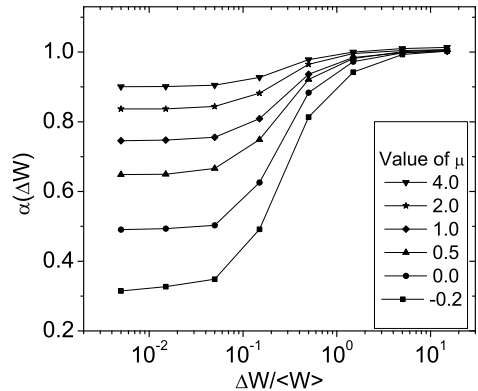}}
\caption{Measured values of $\alpha$ as a function of the relative
strength of driving force ($\Delta W/\ev{W}$) for several fixed values
of $\mu$ ($N=2000$ nodes, $W=200$ walkers, $T=100$ steps/day, averaged
for $D=10^6$ days). Although the internal value ($\Delta W / \ev{W}
\ll 1$) varies, all systems display the universal behavior $\alpha =
1$ in the exogenous limit ($\Delta W/\ev{W} \gg 1$).}
\label{fig:rwm_internal_exponent_driven}
\end{figure}

Next, in order to analyze the behavior of the system under the
influence of an external drive, we allowed day to day changes in the
number of walkers $W$. Following Ref. \cite{barabasi.fluct}, we
introduced $W(t)=\ev{W}+\Delta W(t)$, where $\Delta W(t)$ is Gaussian
white noise with standard deviation $\Delta W$ \cite{foot3}. This acts
as an external driving force and contributes to fluctuations. It is
known, that increasing $\Delta W$ toward the strongly driven limit
($\Delta W / \ev{W} \gg 1$), any system displays a crossover to
$\alpha = 1$ as a sign of the growing dominance of exogenous
behavior. We used the above set of parameters and varied $\Delta
W/\ev{W} = 5\cdot10^{-3},\dots,15$. For all values of the internal
exponent \eqref{eq:alpha} we recovered this expected tendency, as
shown in Fig. \ref{fig:rwm_internal_exponent_driven}. Note that the
intermediate values above that given by \eqref{eq:alpha} but below $1$
are effective exponents, actual scaling breaks down due to the
crossover between them.

This approach can be reversed. Driving hides the microscopic dynamics,
because all systems display the universal value $\alpha=1$. However,
if it is possible to measure the internal exponent $\alpha(\Delta W /
\ev{W} \ll 1)$, one can decide about the presence of impact
inhomogeneity. It has been found \cite{barabasi.fluct}, that for the
hardware level Internet $\alpha=1/2$. In this case $f_i(t)$ is the
data flow through node $i$ at time $t$, $\ev{V(i)}$ is the mean size
of the passing data packets and $\ev{N(i)}$ is their mean number per
unit time. This shows, that across the nodes of this system (routers)
only the number of packets varies, but their size does not ($\mu = 
0$). This homogeneous dynamics can be expected: As the same packets
pass many computers, the mean of their sizes can well be independent
of node degree.

Although not readily represented as a network, a similar analysis can
be carried out on stock market data \cite{eisler.coming}. Here,
$f_i(t)$ is the value of stocks of the company $i$ bought/sold at time
$t$, $\ev{N(i)}$ is the mean number of transactions per unit time, and
$V(i)$ is the value of stocks exchanged in a single trade. The role of
degree (node size) is taken by company capitalization. In this case, it has been
found that $\nu\approx 0.39$ and $\mu\approx 0.44$
\cite{zumbach}. This is direct evidence for the existence of scaling
proposed in \eqref{eq:vi}. Accordingly, $\alpha$ has the non-trivial
value $0.72$ \cite{eisler.non-universality}, due to the presence of
inhomogeneous impacts.

If the dynamics that generates the activities $f_i(t)$ is much slower
than the method used to record them, one can observe the single events
at each node. This happens, e.g., if we track each walker in our 
model. In this case, given the node size distribution $k(i)$, it is 
straightforward to measure $\mu$ and $\nu$ directly from $f_i(t)$ by 
their definitions. Again, this has been possible for stock markets 
\cite{zumbach}, because there all individual trades are documented as 
so called tick-by-tick data. Other possibilities could be distributed 
computing or telephone networks, where events take a longer time, 
while logs of activity can be written instantly. It would be 
interesting to check the validity of our assumptions in these networks 
too.

The partial support of the Center for Applied Mathematics and 
Computational Physics of the BUTE is acknowledged.


\begin{thebibliography}{99}
\bibitem{foot1} The theory itself is valid for any type of network. While our assumptions are independent of scale freedom, it ensures vertices with a broad range of degrees and so of $\ev{f}$. This is necessary for the observation of scaling.
\bibitem{foot2} To correct for bias coming from that there are many low activity nodes, whereas only a few highly active nodes, it is straightforward to take equal size bins in $\log \ev{f_i}$ and average $\log \sigma_i$ for points in the same bin. This way, equal weight is given to all regimes in $\ev{f_i}$. This method was used throughout the paper.
\bibitem{foot3}
Of course the number of walkers $W(t)$ is always non-negative. Thus when $\ev{W}+\Delta W(t)<0$, we define $W(t)$ = 0.
\end{thebibliography}
\end{document}